\documentstyle[11pt,aaspp4]{article}

\begin{document}
\title{NGC~2915 -- A Galaxy with a Dark or Faded Massive Disk?}

\author{
A.~C.\ Quillen\altaffilmark{1}$^,$\altaffilmark{2} 
}
\altaffiltext{1}{University of Arizona, Steward Observatory, Tucson, AZ 85721}
\altaffiltext{2}{E-mail: aquillen@as.arizona.edu}

\def\spose#1{\hbox to 0pt{#1\hss}}
\def\lta{\mathrel{\spose{\lower 3pt\hbox{$\mathchar"218$}}
     \raise 2.0pt\hbox{$\mathchar"13C$}}}
\def\gta{\mathrel{\spose{\lower 3pt\hbox{$\mathchar"218$}}
     \raise 2.0pt\hbox{$\mathchar"13E$}}}

\begin{abstract}
NGC~2915 is a remarkable blue compact dwarf galaxy which contains an
inner \ion{H}{1} bar and strong spiral structure in its \ion{H}{1}
disk extending well beyond its optical disk.
%
We propose here that NGC~2915 has a dark or faint (not yet observed) 
massive disk component.  
By massive we mean of the same order as the \ion{H}{1} gas mass which
ranges from 0.5 -- 10$ M _\odot$/pc$^2$.  We present three
different dynamical arguments:

1) The \ion{H}{1} bar observed in the center is not massive enough to
produce the observed non-circular velocities.  A moderately massive
additional component if it produced shocks in the neutral gas could
cause the observed strong deviations from circular motion.  This bar
would then have mediated the starburst and be responsible for the
\ion{H}{1} bar which could then be similar to the twin peaks observed
in CO in nuclear regions of other barred galaxies.

2) The \ion{H}{1} gas disk by itself is stable to spiral density
waves, in contradiction to the fact that there is spiral structure observed.  
For spiral density waves to exist a more massive disk is
required.

3) Strong spiral structure observed in the outer parts of the galaxy
probably requires forcing greater than can be caused by the neutral
gas density alone.

The additional mass requirements from these three dynamical
arguments depend upon the distance of the
galaxy in the same way.  If the galaxy were at a distance twice that
estimated (10 Mpc instead of 5 Mpc), the additional mass required
would be negligible.  
Virgocentric models show that if the galaxy were this distant 
its radial velocity would be 250-500 km/s below its local bulk flow value.
Local flow models suggest that this is unlikely, however they
also do not exclude it at high confidence levels.

We discuss two possible constituents for the conjectured additional matter,
a molecular component and a stellar component.  The
possible molecular component should be easily detectable in CO
emission.  The possible stellar component also should be easily
detectable with deeper red optical band imaging for moderate mass-to-light
ratios (1-3 in $R$ band).  We favor a possible stellar component because 
late type galaxies galaxies commonly do not have large molecular gas fractions. 

Detection of a quiescent or faint stellar disk population 
would be an exciting prospect since it would be strong 
evidence for a previous epoch
of star formation in a previously low surface brightness galaxy.
If this possibility were confirmed NGC~2915 would be an example
of a galaxy which has faded and so could have been 
one of the faint blue galaxy population observed at moderate redshift.

\end{abstract}


\section {Introduction}

NGC~2915 has optical properties of a weak Blue Compact Dwarf (BCD)
(\cite{meu94}, hereafter MMC)
whereas its \ion{H}{1} morphology and global profiles (\cite{meu96}, hereafter MCBF) are those of an Sd-Sm
disk galaxy (\cite{rob94}; \cite{sho78}).
Remarkably MCBF found that this galaxy
contains an inner \ion{H}{1} bar and strong spiral structure in its 
neutral gas disk
which extends far beyond its optical disk -- 5 times past its Holmberg radius 
(the radius, here $r=1.9'$, at which the $B$ band surface brightness is 
26.6 mag arcsec$^{-2}$).  
Recently \cite{qui97} proposed that the presence of strong spiral structure 
in the neutral gas could be used to place limits on the disk mass of a
low surface brightness galaxy.   In particular they found
that the stellar mass surface density in the two massive
low surface brightness galaxies Malin 2 and UGC 6614 
was of the same order as that in the atomic gas.
In this paper we consider the disk mass in non-axisymmetric
bar and spiral arm structures required 
to produce the strong spiral and bar structures observed in the \ion{H}{1} gas
distribution and velocity field in NGC~2915.

Various works have used perturbations from non-circular motion to place limits 
on the non-axisymmetric component of matter that is not directly visible.
For example by modeling the velocity field of IC 2006 \cite{frx94} found that the
halo of this galaxy was very close to axisymmetric in the plane of the galaxy.
Models for the velocity field of the polar ring galaxy NGC~4650A, on the other 
hand, required a strongly flattened halo (\cite{sac94}).
\cite{eng90} found that the bar mass of NGC~1073 was not sufficiently
strong to produce the observed non-circular gas motions, and that 
an additional elliptical massive disk component was required.
In this paper we consider the possibility of similar additional 
disk components in NGC~2915.

We consider three different approaches on placing limits on the disk mass in 
NGC~2915.
In \S 2 we consider the mass in a bar-shaped disk component sufficiently massive
to produce the observed non-circular motions and strong gas response
in the central regions of the galaxy.
We then (in \S 3) discuss the disk mass required 
for the disk to be unstable enough to support the observed spiral structure.
This is intimately related to the result of MCBF 
that the galaxy did not obey the critical star formation
threshold gas density commonly observed in spiral galaxies (\cite{ken89}).
Thirdly (\S 4) we consider the mass in a non-axisymmetric spiral component
required to produce shocks in the \ion{H}{1} consistent with the
strong spiral structure observed in the \ion{H}{1} gas distribution at large radii.
This estimate is similar to that used in \cite{qui97} to derive lower
limits for the disk mass-to-light ratio in the low surface brightness galaxies.
The limits we place on the disk mass depend on the distance
to NGC~2915 so wherever possible we have
included a distance dependent correction factor.
MMC estimated a distance of $D= 5.3\pm 1.6$ Mpc, to which
we scale all our relations.


In \S 2-4 we find that with our assumed distance 
a more massive disk 
than observed in \ion{H}{1} is required to be consistent with the observations.
Since a distance twice that assumed would make
the additional mass required negligible,
in \S 5 we the discuss the possibility that galaxy is more distant.
We then consider two possible constituents for the conjectured additional
massive disk component: a molecular component and a stellar component.  
Prospects for detection of both possible constituents are then explored.  
A summary and discussion follows.

\section{The Mass of the Bar}

In this section we consider the mass in a non-axisymmetric bar structure 
required to be consistent with the \ion{H}{1} morphology and velocity field. 
In Figure 1 we show an $I$ band image and the \ion{H}{1} column
density map and velocity contours (from MCBF).
It was noted in MCBF that within a diameter of
11 kpc or 7' the velocity field shows the characteristic twist of
non-circular motion driven by a bar.  
Tilted ring fits were unable to account for the twist in the velocity
field which contained residuals of $\sim 10$ km/s along the minor axis 
over the region $200'' < r < 450''$.     For gas undergoing circular motion
the zero velocity line of nodes is always perpendicular to the maximal
velocity line of nodes in the velocity field.  However these lines
of nodes are not perpendicular for gas in non-circular or oval orbits.
The failure of the tilted ring fits along the minor axis is  strong
evidence for non-circular motion.  MCBF also observed that both the 
\ion{H}{1} gas distribution and the optical isophotes 
are elongated in the central region.  The \ion{H}{1} gas distribution shows 
a prominent central bar about 2.5' (3.9 kpc) long and 
the optical isophotes, while only slightly elongated, do have 
major axes roughly aligned with the \ion{H}{1} bar major axis.

The magnitude of the velocity
deviations from circular motion within are significant (at least $10$ km/s).   
This is in striking contrast to the contribution to the circular 
velocity predicted from the light (displayed in Fig. 16 of MCBF)
which reaches a maximum of only 30 km/s at less than 1kpc and drops
quickly, falling to 15km/s at a radius of $r=5$ kpc or $190''$, 
and the circular velocity contribution 
predicted from the neutral gas, which is flat
at 15--18 km/s for $r= 1$ -- $15$ kpc ($38''$--$580''$).
We ask here how can these two visible components be massive enough
to cause the large observed deviations from circular motion?

\subsection{The Strength of the \ion{H}{1} Bar}

MCBF found that that even in the bar region, the contribution
to the circular motion from the stars (traced in optical images) and from the 
gas was small compared to the circular velocity.  As a result
they found that the dark matter component required to fit the \ion{H}{1} 
derived rotation curve dominated at all radii.
Since the stars and gas contribute only negligibly to the circular velocity
the magnitude of the oval non-axisymmetric (or bar) potential components resulting 
from the gas and stars must be small compared to the 
axisymmetric (presumably dark matter dominated) component.
To confirm this we have computed the $m=2$ Fourier component of the gravitational
potential, $\Phi_2$, (displayed in Fig. 2)
generated from the neutral gas distribution 
(corrected for He content from the hydrogen with a factor of 1.33) 
using the technique of \cite{qui94} assuming 
an inclination of $58^\circ$ and a position
angle of $-70^\circ$ (from the titled ring fit to the velocity
field of MCBF outside the bar region). 

We see that the $\Phi_2$ or $m=2$ Fourier components of the 
gravitational potential are only a small
fraction of the axisymmetric component derived from the observed rotation curve 
which is approximately flat at 80km/s past a radius of $200''$ or 5.2 kpc.
The magnitude of $\Phi_2$ is somewhat dependent on the inclination where
a $5^\circ$ lower assumed inclination results in a reduction in the $\Phi_2$
components of $\sim 10\%$.
Using the $m=2$ components of the potential and the rotation curve presented 
in MCBF, we estimate the velocity perturbations for gas in elliptical orbits
in this potential.  The resulting velocity perturbations (computed using Eqs. 6.30
from \cite{B+T}) are also shown in Figure 2.
We have neglected the pressure from the gas in these equations 
which has the opposite sign of the potential terms and so tends to reduce
the magnitude of the velocity residuals.
We have also assumed a bar pattern speed which places the corotation 
radius at $300''$ or 7.8kpc.  This bar rotation rate places the \ion{H}{1} bar
at the Inner Lindblad Resonance (similar to the twin molecular peaks
of \cite{ken92}) and the end of the bar at $\sim 250''$ where the 
$\Phi_2$ component begins to twist or vary in position angle indicating
the onset of spiral structure outside the bar.

As expected (see Fig. 2d) the predicted velocity deviations from circular motion 
are small, much smaller than those observed except directly at resonances.  
Since the deviations from circular motion are detected over range of radius
$200'' < r < 450''$ (and not confined to small regions that could be near resonances)
these predicted velocity deviations are not sufficiently
large to account for the observed non-circular motions.
Varying the bar pattern speed since it only shifts the resonances does not
allow stronger velocity deviations over the range of radius observed.
In fact over much of the region where there is evidence for non-circular motion
the predicted velocity deviations are also substantially smaller than 
the gas velocity 
dispersion of the outer disk (measured by MCBF $\sigma \sim 8$ km/s).   
If the atomic gas were the only non-axisymmetric mass 
component, the gas response should therefore be smooth, and non shocked
over a large range of radius. 
The gas would be in nearly circular orbits and so should not exhibit 
strong observed departures from circular motion.  Consequently it
is also improbable that the gas would be confined to the prominent \ion{H}{1} bar 
observed.  We therefore find that it is unlikely that the twists in the 
velocity field in the central region of the galaxy 
are caused solely by the mass in the \ion{H}{1} bar itself.  The optical component
is only slightly elongated and only contributes significant (in mass)
at small radii (within $\sim 100''$)
and so is also not sufficiently massive or elongated 
to produce the observed non-circular motions.

\subsection{A Possible Additional Massive Bar Component}

We now consider the possibility that non-circular motions are caused by a massive
bar that has not been detected.  If such a bar were to exist in NGC~2915 it would
be a natural explanation for a variety of observations.  The bar could have caused
gas inflow which in turn produced the 
the blue compact nucleus, which is actively 
forming stars and contains a relatively young stellar population (MMC; \cite{mar95}).  
MCBF noted that with higher spatial resolution the \ion{H}{1} bar is
resolved into two clouds.  
This suggests that the \ion{H}{1} bar might be similar
to those observed in molecular gas in barred galaxies, like  
the ``twin peaks'' seen in CO of \cite{ken92}, which
are suspected to be located at the bar's Inner Lindblad Resonance.   
This would place the bar corotation radius roughly at $300''$ which is 
roughly where
the velocity twist reverses in direction as expected at the radius of corotation.
The double peaked velocity profiles observed in the \ion{H}{1} bar (MCBF)
could then be interpreted in terms of a velocity jump across a shock region.

Strong non-circular motions in the gas can be caused by 
a modest non-axisymmetric
or bar-like perturbation to the gravitational potential if the gas is undergoing
shocks.   This was stressed in early studies such as \cite{rob79} who noted
that only a 10\% oval perturbation to the potential (which might result from 
a $\sim 30\%$ perturbation in the density) 
was sufficiently strong in normal galaxies 
to cause strong non-circular motion in the gas response
because shocks were produced in the gas.  

We can make a rough estimate of the mass needed to produce shocks in the gas
by considering the size of a $\Phi_2$ component required to raise the
velocity perturbations to the magnitude of gas velocity dispersion
observed in the outer disk
$\sigma \sim 8$ km/s (MCBF).  When the velocity perturbations are of this magnitude
the gas flow can be supersonic and shocks can occur.  
This in turn allows  large departures from circular motion as well as
strong gas density contrasts.  In Fig. 2d we show that
for the \ion{H}{1} mass distribution the velocity perturbations 
are $\sim 2$ km/s ($D / 5.3$ Mpc)  over much of the bar region.
For these velocity perturbations to be of the same order
as the gas velocity dispersion a mass 
approximately four times the neutral gas density is required to produce shocks.
The extended neutral gas surface density 
in the \ion{H}{1} bar region ranges from $1.5$ -- $4 M_\odot /{\rm pc}^2$
for $2' < r < 6'$.
We therefor estimate that the additional mass required is 
$\sim 6$--$16 M_\odot/{\rm pc}^2$ (5.3 Mpc /$D$) in the bar region.
For an extended bar with major axis profile that is roughly of
constant surface brightness (similar to those seen in barred
galaxies and numerical simulations) a surface density of
$\sim 6 M_\odot/{\rm pc}^2$ would be required.

\section{Instability to Spiral Structure}

It was noted by MCBF that the gas density is a factor
of about 3 (range of 2-9) below the critical density, $\Sigma_{crit}$,
for star formation (\cite{ken89}), despite the star formation activity in
the central regions.   This critical density is defined in \cite{ken89} as
\begin{equation}
\Sigma_{crit} \equiv {\alpha \kappa \sigma \over 3.36 G}
\end{equation}
where $\alpha=0.7$ was determined empirically.
This critical density is intimately related to the Toomre stability parameter 
\begin{equation}
Q \equiv {\kappa \sigma \over 3.36 G \Sigma} = {\Sigma_{crit} \over \Sigma \alpha}
\end{equation}
(e.g. see \cite{B+T}) where $\kappa$ is the epicyclic frequency 
and $\Sigma$ is its mass surface density.
We note that when $Q > 2$ amplification
processes such as the swing amplifier are inefficient and the disk is unresponsive
to tidal perturbations which could excite spiral density waves in a more
unstable disk (see \cite{B+T} and references theirin).
It is therefore unlikely for a disk with $Q>2$ to show spiral structure.
This implies that a disk with gas density below the critical gas density
of \cite{ken89}
is also unlikely to show spiral structure assuming there is no other
disk mass component.  (Note that $Q\sim 1.4 $ for $\Sigma_{crit} /\Sigma = 1$.)
Consequently if a gas disk is well below the critical gas density
and yet shows spiral structure, a natural explanation is that there
is another massive component in the disk
(see \cite{jog84} for instability in a two fluid disk).

The ratio of $\Sigma_{crit}/\Sigma$ found by MCBF implies that the 
Toomre parameter Q is on average 4.2 (range of 3 -- 13) 
assuming a single fluid \ion{H}{1} disk.
(Multiply these values by a factor (5.3 Mpc /$D$) 
if a different distance to NGC~2915 is assumed).
These large values for $Q$ would suggest that the gas disk should not
show any spiral structure, in contradiction with the observations.
The low gas densities of NGC~2915 coupled with its strong spiral
structure are strong evidence for the existence of an additional
massive disk component.

To estimate the mass of this additional massive disk component 
some assumption must be made about its velocity dispersion $\sigma_d$.  
For a stellar component of the same dispersion as the gas $\sigma \sim 8$ km/s
for $Q<2$ a disk mass of at least the mass surface density of the \ion{H}{1} disk is required.
If the additional mass component is molecular, then it could
have a lower velocity dispersion and so a lower mass density would
be required.  For example a molecular component with
a dispersion half that of the \ion{H}{1} a disk surface density approximately half
of that of the neutral gas would be required for the disk to be unstable
enough to support the observed spiral structure.
Using the neutral gas surface density measured from the \ion{H}{1}
observations, we estimate that the additional mass component 
should have mass surface density approximately
$0.5 - 2 M_\odot/{\rm pc}^2 \left({5.3 {\rm Mpc} \over D}\right) 
\left({\sigma_d \over 8 {\rm km/s}}\right)$ for $r>300''$ outside the bar region.

\section{Critical Spiral Forcing Required to Produce Shocks }

The large density variations observed in outer spiral arms 
of NGC~2915 are evidence for shocks in the ISM induced by a spiral gravitational potential. 
Indeed the high arm/interarm density
contrast of \ion{H}{1} observed in galaxies such as M81 and M51 is one
of the major predictions of the spiral density wave theory (e.g. \cite{B+T} 
on the Lin-Shu hypothesis).  In this
section we consider how much mass is required in the form of spiral
structure to drive shocks in the gas that would be consistent with the
$\sim$2:1 arm/interarm density contrasts observed in the \ion{H}{1} of NGC 2915.

A critical forcing parameter to produce shocks or large density contrasts in the
ISM was explored by \cite{rob69} and \cite{shu73}.  
These authors considered the role of $F$, the spiral 
gravitational force expressed as percentage of the axisymmetric force which
can be written in terms of the density variation observed
in the spiral structure (\cite{qui97}) as 
\begin {equation}
F  = {2 \pi G \Sigma_2  r \over v_c^2}
\end {equation}
using the WKB or tight winding and thin disk approximations.
Here $v_c$ is the circular velocity. 
The forcing parameter depends on the size of the mass density variations
in the spiral arms or on
$\Sigma_2$, the magnitude of the $m=2$ Fourier components of the density 
at a radius of $r$.
$F \gta  2\%$ is generally required for shocks to form 
(\cite{rob69}, \cite{shu73}, see also
\cite{too77}).  However this value was estimated
in a Milky Way sized galaxy with rotational velocity of 200 km/s
and gas velocity dispersion of $\sim 8$km/s.  Since the forcing
required to produce shocks is expected to depend on the ratio of
the sound speed to the circular velocity, we expect that that
forcing twice as large or $F \gta 4\%$ could be 
required to produced shocks in NGC~2915 where the rotational 
velocity is less than half that of the Milky Way.  For a short review of forcing 
requirements see \cite{too77} or \cite{qui97}. 

In Table 1 we have estimated the forcing parameter $F$ using
the above equation and the neutral gas density variations (derived
from the \ion{H}{1}) for radii outside of the bar region.
Table 1 shows that the forcing parameter $F$ is less than 2\% in the outer parts
of NGC~2915 and so is probably insufficiently strong to cause shocks in the ISM and
the large neutral gas density contrasts observed.
For a forcing of $F \gta 2\%$  an additional disk mass component 
is required of approximately the same mass density as the neutral gas,
if this additional mass component has the same density variations
observed in \ion{H}{1}.
The mass of this component would then be 
$0.5 - 2 M_\odot/{\rm pc}^2 
\left({5.3 {\rm Mpc} \over D}\right) 
\left({\Sigma_d \over \Sigma_{2,HI} {\rm km/s}}\right)$  
where the mass density required depends 
its azimuthal variations $\Sigma_d$ or on the magnitude
of its $m=2$ Fourier component.

\section{ A Summary of Additional Disk Mass Estimates} 

Here we summarize the excess disk mass requirements 
which we have estimated in the previous three sections.

1) Mass required in a bar component strong enough to 
cause shocks in the gas resulting in large scale non-circular velocities 
consistent with the observed velocities and the \ion{H}{1} bar itself:
$6-16 M_\odot/{\rm pc}^2 ~ (5.3 {\rm Mpc} /D)$ for $r<300''$.

2) From the stability of the disk an excess surface density
of the same order as the \ion{H}{1} surface density:
$0.5 - 2 M_\odot/{\rm pc}^2 \left({5.3 {\rm Mpc} \over D}\right) 
\left({\sigma_d \over 8 {\rm km/s}}\right)$  
for $r>300''$
where the mass density depends upon its the velocity dispersion,  $\sigma_d$. 

3) From the spiral forcing required in the outer disk 
an excess surface density 
of the same order as the \ion{H}{1} surface density: 
$0.5 - 2 M_\odot/{\rm pc}^2 \left({5.3 {\rm Mpc} \over D}\right) 
\left({\Sigma_d \over \Sigma_{2,HI} {\rm km/s}}\right)$  
for $r>300''$
where the mass density depends on its azimuthal variations, $\Sigma_d$.

We note that the mass surface densities we estimate from the above arguements are 
only approximate.  Better estimates could be made with gas simulations
such as those in \cite{low94} for spiral structure and in
\cite{eng90} in barred galaxies.

\subsection{Could NGC~2915 Be More Distant?}

We note from the above summary that if the galaxy is twice as distant
as 5.3 Mpc that the additional mass required would be negligible.
How likely is this?   Here we review distance estimates to the galaxy.
MMC estimated the distance to NGC~2915 of $5.3\pm 1.6$ Mpc, 
by scaling the properties of stars resolved in its compact
core to those of the galaxy NGC~5253 which has a similar
morphology.  The distance of NGC~5253 has been
accurately measured using Cepheid variables (\cite{san94}).
As in the compact nucleus  of NGC~5253, MMC found that 
many of the resolved objects in NGC~2915 are most likely individual 
massive stars.  For comparison,
based on its radial velocity ($v_{heliocentric} = 468 \pm 5$ km/s)
and a  linear virgocentric inflow model, 
\cite{sch92} estimate a distance of 4.1 Mpc to NGC~2915 consistent with
the estimate of MMC (5.3 Mpc).  Alternatively, the linear virgocentric model adopted
in \cite{mar95} gives a distance of 2.9 Mpc.  Using
the observed radial velocity (and no virgocentric model) 
a Hubble constant of $H_0 = 75$ km s$^{-1}$ Mpc$^{-1}$ 
gives a distance of 6.2 Mpc.

If NGC~2915 is actually at 10 Mpc  then it would have a substantial
peculiar velocity compared to the local bulk flow.  Using
a virgocentric model with a distance to Virgo of 15.9 Mpc (based on
($H_0 = 75$  km s$^{-1}$ Mpc$^{-1}$), and local inflow 
velocity of 220 km/s we estimate that
if the true galaxy distance were 10 Mpc, its difference in radial velocity 
from the bulk flow would be $\sim 460$ km/s.  
This is large compared to
the RMS variations in peculiar velocity from the bulk 
flow estimated by \cite{aar82}  ($\sim 150$ km/s)
for galaxies within the local supercluster and that 
($\sim 100$ km/s, \cite{san72}; \cite{fab88}) 
estimated for galaxies moving with the Local Group or with
radial velocities less than 700 km/s. 
Using these standard deviations NGC~2915 would be 3-4 sigma
away from the bulk flow.
However for $H_0 =56$ and using
a virgocentric model consistent with this value, NGC~2915's  difference
in radial velocity would be only 260 km/s which is less than 2 sigma away
from the local bulk flow.
We also note that \cite{fab88} found that the standard deviation
in peculiar velocity from the bulk flow
is substantially larger, $\sim 300$ km/s, for galaxies outside the Local
Group flow, so if the galaxy is more distant a large deviation 
in peculiar velocity is not necessarily unlikely.

These estimates show that if NGC~2915 is at 10Mpc its radial
velocity would be 250--500 km/s below its local bulk flow value.
Local supercluster flow models suggest that this is unlikely, however they
also do not exclude this distance with high confidence levels.

\section{The Possible Constituents of the Additional Disk Component}

There are two likely possibilities for our conjectured additional 
mass:  a molecular component or a faint stellar component.
Here we discuss both possibilities. 

\subsection{A Possible Molecular Component}

A molecular component in the outer disk could have a low velocity
dispersion and a large spiral arm  density contrast, so that
the mass required for such a component in the outer disk could be
approximately half that of the neutral gas disk.  
However a high gas mass in the form of a central bar would still be required,
(although then it would then be quite mysterious that such molecular component
is not vigorously forming stars).
With a molecular to atomic mass ratio of $M(H_2)/ M(HI)\sim  0.5$,
and a total neutral gas mass of $\sim 10^9 M_\odot$ (MCBF), 
we estimate that
the total molecular gas mass would be $\sim 10^5 M_\odot$.
Such a molecular gas mass should be easily detectable in emission in CO. 
We note that if the metallicity ($\lta 0.6 Z_\odot$, MCBF)
is not extremely low, the CO to H$_2$ conversion factor should 
not be high compared to the Milky Way.

Late-type galaxies have lower molecular gas fractions
than early-type spiral galaxies.  For example,
the Sd/Sm galaxies shown in \cite{you91} have a mean $M(H_2)/M(HI) = 0.1$ 
(\cite{you89}, \cite{thr89}) with scatter at the level
of 0.2.  This suggests that a massive molecular component is possible,
though unlikely, particularly since the large concentrated gas masses
required would be expected to be actively forming stars in the bar and outer
disk (as in \cite{vog88}).
The survey of Blue Compact Dwarfs of \cite{sag92}
found that these galaxies typically had low molecular to atomic 
gas fractions.  It is therefore unlikely that a significant percent of
the disk mass in NGC~2915 is molecular gas.


\subsection{A Possible Stellar Component}

The conjectured stellar component must have surface brightness
lower than that seen in the images presented in MMC and MCBF.
The $B$ band image of MMC detected the galaxy out to 27 mag/arcsec$^2$
at $r\sim 120''$,
whereas their $R$ band image had a detection limit at 24.5 mag/arcsec$^2$
and only detected the galaxy at smaller radii 
presumably because of its red color (the outermost measured points have 
$B-R \sim 1.6$).
The $I$ band image of MCBF is unfortunately uncalibrated but is 
similar to the $R$ band image in appearance and depth.
The outer disk must therefor have have surface brightnesses greater than 
27 and 24.5 mag/arcsec$^2$, in $B$ and $R$ bands respectively.
For comparison a $B$ and $R$ band surface brightness of $27$ and  $26$ 
mag/arcsec$^{2}$ with $M/L_B = 1$, $M/L_R =1$ (in solar units) 
are both equivalent to a surface density of $1.0 M_\odot /{\rm pc}^2$.

For an extended bar with surface density of $\sim 6 M_\odot/{\rm pc}^2$
detected at a level fainter than the limiting surface brightnesses 27
and 24.5 mag/arcsec$^2$ in $B$ and $R$ bands, would have 
mass-to-light ratios
$M/L \gta$ 6 and 1.5 in $B$ and $R$ bands respectively,
For an outer disk with 
surface density of $\sim 1 M_\odot /{\rm pc}^2$ mass-to-light ratios
$M/L \gta 1$ and 0.25 are required in $B$ and $R$ bands respectively
for a disk fainter than the above limits.

We now compare these mass-to-light ratio values with those predicted
with population synthesis models and maximal disk fits to rotation
curves.  The reddest, and most abrupt exponential burst models of \cite{ken94}
at 10 Gyr have $M/L_B = 4-8$ and  $M/L_R = 3-5$.
The single burst models of \cite{wor94} have approximately the same 
mass-to-light ratios for a moderate metallicity stellar population 
of age 12 Gyr.
This is to be compared to maximal disk models for normal spiral 
galaxies which have $M/L_R = 1$--$7$ (\cite{ken87a}, \cite{ken87b}),
consistent with the hypothesis that the dark matter contribution 
is negligible in the central few scale lengths of these galaxies.

The above limits show that the conjectured stellar disk 
could have mass-to-light ratios of an old stellar population.   
In this case we expect expect $M/L_R \gta 3$ otherwise
the disk would have extremely red colors (the disk is faint at $B$ band).
For an old stellar population (see above) with $M/L_B \sim 6$ and 
$M/L_R \sim 3-4$ and the reddish colors of $B-R \sim 1.5$,  
the disk should be detectable with deeper optical imaging
in a red band ($R$ or $I$) particularly in the bar region.
The mass-to-light
ratios are required to be relatively high in the bar because of the large
bar mass we estimate.  
The outer disk, however, would be difficult to detect unless it had lower 
mass-to-light ratios.

\section{Summary and Discussion}

We have presented three arguements that suggest there is a dark or faint
massive disk in NGC~2915.  By massive we mean with 
mass surface density of the same order as that in 
the \ion{H}{1} gas, or a few $\sim M _\odot$/pc$^2$.
The mass surface densities estimated are listed in the previous section.

1) The \ion{H}{1} bar observed in the center is not massive enough to produce
the observed non-circular velocities.
A moderately massive additional component if it produced shocks 
in the neutral gas
could cause the observed strong deviations from circular motion.
This bar would then have mediated the starburst and be responsible 
for the \ion{H}{1}
bar which could then be similar to the twin peaks observed in CO in 
nuclear regions of other barred galaxies.
Double peaked \ion{H}{1} velocity profiles detected at these \ion{H}{1} peaks
(MCBF) could then be naturally interpreted in terms of velocities 
on either side of a galactic shock.

2) The outer \ion{H}{1} gas disk by itself is stable to spiral density waves,
in contradition to the fact that spiral structure is observed.  
For spiral density waves to exist a more massive disk is required.

3) Strong spiral structure observed in the outer parts of the galaxy probably
requires spiral graviational forcing greater than can be caused 
by the neutral gas density itself.

The additional mass requirements all depend upon the distance of the galaxy
in the same way.  If the galaxy were at a distance twice that assumed here (10 Mpc
instead of 5 Mpc), the additional disk mass required would be negligible.
Virgocentric models show that if the galaxy were this distant
its radial velocity would be 250-500 km/s below its local bulk flow value.
Local flow models suggest that this is unlikely, however they
also do not exclude it at high confidence levels.
A more precise measurement of the distance to NGC~2915 
is required to confirm the presence of a dark or faint additional disk component.

We discuss two possible constituents for the 
conjectured additional disk mass component, a molecular
component and a stellar component.  
The possible molecular component should be easily detectable in CO emission.
A possible stellar component could be an old stellar population,
in which case 
it should be detectable with deeper $R$ or $I$ band optical imaging,
particularly in the bar region  
for $M/L_R \sim 3--4$ which
would give the galaxy red colors consistent with the limiting
surface brightness observed in $B$ band and $M/L_B \gta 6$ 
needed to give sufficient mass to the bar.
We favor a possible stellar component because late-type galaxies 
usually have low molecular to atomic gas fractions, and 
it would be difficult to explain why such a molecular gas component 
is not actively forming stars, particularly in the bar.
no visible evidence of star formation in the outer disk or bar.

We note that if further observations 
of the galaxy
confirm its near solar metallicity ($\lta 0.6 Z_\odot$) then this would
imply that a significant fraction of its gas had undergone previous enrichment.
This would be consistent with the proposed additional stellar disk component.

If the existence of an additional massive disk component is confirmed
this would be evidence for a previous epoch of star formation
in a galaxy which was until quite recently 
(before its present Blue Compact Dwarf phase) a low surface brightness galaxy.  
A population of red low surface brightness galaxies  has been
recently discovered (\cite{one97a}, \cite{one97b}) and NGC~2915 could
similar to one of these galaxies, but with a 
newly formed blue compact core.
NGC~2915 would then have a faded disk and could have been 
one of the faint blue galaxy population 
observed at moderate redshift (\cite{bro88}, \cite{efs91}, \cite{col90}).   

Low surface brightness galaxies 
are natural sites for a search and study
of previous epochs of star formation since the low levels of recent star
formation facilitates observing an older dimmer stellar population.
We note that only an undisturbed quiescent field galaxy 
could maintain a cold stellar disk, without heating its 
stars and dispersing them into a thick disk or halo
population which would then be unresponsive to spiral structure or bar
formation.  

If the disk in NGC~2915 contains an additional mass component 
then the halo mass required to fit the rotation curve could be much reduced.
NGC~2915 would then be composed of populations with very different
mass-to-light ratios and distributions.
The resulting disk population would exhibit large variations in 
its mass-to-light ratio as a function of radius.    
This in turn would suggest that
we reexamine rotation curve fits which predict halo parameters
by assuming constant disk and bulge mass-to-light ratios in late-type,
dwarf and low surface brightness galaxies.

\acknowledgments

I thank G. Meurer for working with me on this project and providing
me with his data to do the study outlined in this paper.
I thank him for many helpful discussions.
We also acknowledge helpful discussions and correspondence with R.\
Kennicutt, D.\ Hunter, T.\ Pickering, C.~E.\ Walker, L.\ van Zee, 
C.\ Impey, R.\ Allen.

\clearpage


\newpage

\begin{figure*}
\caption[junk]{ 
a) \ion{H}{1} intensity (grayscale).  Contours are shown
with lowest contour at $2 M_\odot/{\rm pc}^2$ and a difference
between contours equal to this value.  At higher resolution
the bar resolves into two peaks.
The beamsize for the \ion{H}{1} observations is shown on the lower
left corner.  Note that the surface density contrast is
high with an arm/interarm contrast of $\sim$2:1.  The high gas
density contrast is a consequence of shocks in the ISM and one of the
predictions of strong spiral density waves. 
These data are from MCBF.
b) $I$ band image of NGC 2915 shown as grayscale. 
These data are from MMC.  The \ion{H}{1} disk extends 5 times past its
Holmberg radius (radius $r=1.9'$ at which the $B$ band surface
brightness is 26.6mag arcsec$^{-2}$).
c) \ion{H}{1} velocity contours with a spacing of 10 km/s with the lowest
contour on the lower left at 400 km/s.   Note the twist in the velocity
field probably caused by bar induced non-circular motions.
These data are from MCBF.
\label{fig:fig1} }
\end{figure*}

\begin{figure*}
\caption[junk]{ 
a) Axisymmetric or azimuthally averaged 
component of the gravitational potential, 
$\Phi_0$, prediction from the \ion{H}{1} or neutral gas distribution.
b) Contribution to the circular velocity using $\Phi_0$ derived from the 
neutral gas density.
The shape of this curve is somewhat different from that
shown in MCBF because the azimuthal average was taken after the potential
was calculated from the density instead of before.
c) The $m=2$ Fourier components of the graviational potential.  The solid and dotted lines
represent the sine and cosine components for an angle of zero aligned with the 
\ion{H}{1} bar.
d) Velocity perturbations predicted from the $m=2$ potential components presented
in c) assuming the rotation curve derived in MCBF.   The solid and dotted lines
are the radial and tangential velocity components. For further information see
\S 2.  There are peaks in the velocity perturbations at the Inner Lindblad
resonance ($\sim 80''$) and Outer Lindblad Resonance ($\sim 500''$).  Note
however that at most radii, the velocity perturbations are negligible 
($lta 2$ km/s) and so
are not large enough to produce the observed departures from circular motion
over the range of radius observed.
If a galaxy distance other than $D = 5.3$ Mpc is preferred 
$\Phi_0$, the $\Phi_2$ components
and the velocity perturbations should be multiplied by ($D/5.3$ Mpc) and 
the circular velocity, shown in b), should be multiplied by the square root of this factor.
\label{fig:fig2} }
\end{figure*}

\newpage

\clearpage
\begin{deluxetable}{cccccc}

\footnotesize
\scriptsize
\tablewidth{200pt}
\tablecaption{Spiral Forcing Strength\tablenotemark{a}\label{tab1}}
\tablehead{
\colhead{$r$} & 
\colhead{$v_c$\tablenotemark{b}} & 
\colhead{$\Sigma_{HI,0}$\tablenotemark{c}  } & 
\colhead{$\Sigma_2/\Sigma_0$\tablenotemark{d}}  & 
\colhead{$F$\tablenotemark{a}} \nl
\colhead{arcsec} & 
\colhead{km/s} & 
\colhead{$M_\odot/{\rm pc}^2$} & 
\colhead{} & 
\colhead{\%} 
}
%
\startdata
300 &  82 & 1.5   &  0.29 & 1.4 \nl
424 &  82 & 1.0   &  0.22 & 1.1 \nl
502 &  82 & 0.8   &  0.20 & 1.0 \nl
557 &  88 & 0.5   &  0.25 & 0.7 \nl
\enddata
\tablenotetext{a}{The spiral forcing, $F$, is expressed as a percentage of the 
axisymmetric force, and is calculated with Eq. 3; see \S 4 for details. 
$F$ should be multiplied by the factor ($D/5.3$ Mpc) if a different
distance to NGC~2915 is desired.}
\tablenotetext{b}{The rotation curve from MCBF.}
\tablenotetext{c}{The azimuthally averaged value of the neutral gas surface density.}
\tablenotetext{d}{$m=2$ Fourier component of the HI column density expressed as a 
fraction of the azimuthally averaged value.}
\end{deluxetable}


\begin{thebibliography}{}

\bibitem[Aaronson et al. 1982]{aar82}
Aaronson, M., Huchra, J., Mould, J., Schechter, P.~L., \& Tully, R.~B. \
1982, \apj, 258, 64

\bibitem[Binney \& Tremaine 1987]{B+T}
Binney, J., \& Tremaine, S. \ 1987, Galactic Dynamics,
(Princeton U. Press).

\bibitem[Broadhurst et al. 1988]{bro88}
Broadhurst, T.~J., Ellis, R.~S., \& Shanks, T. \ 1988, \mnras, 235, 827

\bibitem[Colless et al. 1990]{col90}
Colless, M., Ellis, R., Taylor, K. \& Hook, R. \ 
1990, \mnras, 244, 408

\bibitem[England et al. 1990]{eng90}
England, M.~N., Gottesman, S.~T., \& Hunter, J.~H.~,Jr. \ 1990,
\apj, 348, 456

\bibitem[Efstathiou et al. 1991]{efs91}
Efstathiou, G., Bernstein, G., Tyson, K.~A., Katz, N., \& Guhathakurta, P.
\ 1991, \apj, 380, L47

\bibitem[Faber \& Burstein 1988]{fab88}
Faber, S. \& Burstein, D.  \ 1988,
in `Large Scale Motions in the Universe: A Vatican Study Week',
eds. V. C Rubin, \& G. V. Coyne, S. J.,
Princeton University Press, Princeton, NJ, p. 116 

\bibitem[Franx et al.\ 1994]{frx94}
Franx, M., van Gorkom, J. H., \& de Zeeuw, P. T. 1994, \apj, 436, 642

\bibitem[Jog \& Solomon 1984]{jog84}
Jog, C.~H., \& Solomon, P.~M. \ 1984,  \apj, 276, 114

\bibitem[Kenney et al. 1992]{ken92}
Kenney, J.~D.~P., Wilson, C.~D., Scoville, N.~Z., Devereux, N.~A., \&
Young, J.~S. \ 1992, \apj, 395, L79

\bibitem[Kennicutt 1989]{ken89}
Kennicutt, R.~C. \ 1989, \apj, 344, 685

\bibitem[Kennicutt et al.\ 1994]{ken94}
Kennicutt, R.~C., Jr., Tamblyn, P., \& Congdon, C.~E. \
1994, \apj, 435, 22

\bibitem[Kent 1987a]{ken87a}
Kent, S.~M. \ 1987a, \aj, 91, 1301

\bibitem[Kent 1987b]{ken87b}
Kent, S.~M. \ 1987b, \aj, 93, 816

\bibitem[Lowe et al.\ 1994]{low94}
Lowe, S.~A., Roberts, W.~W., Yang, J., Bertin, G., \&  Lin, C.~C. \ 1994,
\apj, 427, 184
 
\bibitem[Marlowe et al. 1995]{mar95}
Marlowe, A.~T., Heckman, T.~M., Wyse, R.~F.~G., \& Schommer, R. \ 
1995, \apj, 438, 563

\bibitem[Meurer et al. 1996]{meu96}
Meurer, G.~R., Carignan, C., Beaulieu, S.~F. \& Freeman, K.~C. 
\ 1996, \aj, 111, 1551 (MCBF)

\bibitem[Meurer et al. 1994]{meu94}
Meurer, G.~R, Mackie, G., \& Carignan, C.
\ 1994, \aj, 107, 2021 (MMC)

\bibitem[O'Neil et al. 1997a]{one97a}
O'Neil, K., Bothun, G. \& Cornell, M. \ 1997a, \aj, in press

\bibitem[O'Neil et al. 1997b]{one97b}
O'Neil, K., Bothun, G., Cornell, M., \& Impey, C. \ 1997b, \aj, submitted 

\bibitem[Quillen et al. 1994]{qui94}
Quillen, A.~C., Frogel, J.~A., \& Gonz\'alez, R.~A. \ 1994,
\apj, 437, 162

\bibitem[Quillen \& Pickering 1997]{qui97}
Quillen, A.~C., \& Pickering, T.~E. \ 1997, \aj, in press, astro-ph/9701159

\bibitem[Roberts \& Haynes 1994]{rob94}
Roberts, M.~S., \& Haynes, M.~P. \  1994, ARA\&A, 59, 19

\bibitem[Roberts et al. 1979]{rob79}
Roberts, W.~W., Jr., Van Albada, G.~D., \& Huntley, J.~M. 
\ 1979, \apj, 233, 67

\bibitem[Roberts 1969]{rob69}
Roberts, W.~W., Jr. \ 1969, \apj, 158, 123

\bibitem[Sandage 1972]{san72}
Sandage, A. \ 1972, \apj, 178, 1

\bibitem[Sandage et al. 1994]{san94}
Sandage, A., Saha, A., Tamman, G.~A., Labhardt, L., Schwengeler, H., 
Panagia, N., \& Macchetto, F.~P. \ 1994, \apj, 432, L13

\bibitem[Sackett et al. 1994]{sac94}
Sackett, P. D., Rix, H., Jarvis, B. J., \& Freeman, K. C.\ 1994, \apj, 436, 629

\bibitem[Sage et al. 1992]{sag92}
Sage, L.~J., Salzer, J.~J., Loose, H.~-H., \& Henkel, C.
\ 1992, A\&A, 265, 19

\bibitem[Schmidt \& Boller 1992]{sch92}
Schmidt, K.-H., \& Boller, T. \ 1992, aNac, 313, 189

\bibitem[Shostak 1978]{sho78}
Shostak, G.~S. \ 1978, A\&A, 68, 321

\bibitem[Thronson et al 1989]{thr89}
Thronson, H., Tacconi. L., Kenney, J., Greenhouse, M., Margulis, M. et al. \ 
1989,  \apj, 344, 747


\bibitem[Young \& Scoville 1991]{you91}
Young, J.~S., \& Scoville, N. Z \ 1991, \araa, 29, 581

\bibitem[Young \& Knezek 1989]{you89}
Young, J.~S., \& Knezek, P. M. \  1989, \apj, 347, L55

\bibitem[Shu et al.\ 1973]{shu73}
Shu, F.~H., Milione, V. \& Roberts, W.~W. \ 1973, \apj, 183, 819

\bibitem[Toomre 1977]{too77}
Toomre, A. 1977, \araa, 15, 437

\bibitem[Vogel et al. 1988]{vog88}
Vogel, S.~N., Kulkarni, S.~R., \& Scoville, N.~A. \ 1988, \nat, 334, 402

\bibitem[Worthey 1994]{wor94}
Worthey, G. \ 1994, \apjs, 95, 107















\end{thebibliography}
\end{document}